\documentclass[a4paper]{spie}  %>>> use this instead for A4 paper
\usepackage[]{graphicx}

\newcommand{\mps}{\mbox{m\,${\rm s}^{-1}\,$}}
\newcommand{\kmps}{\mbox{km\,${\rm s}^{-1}\,$}}
\newcommand{\dps}{\mbox{$^\circ$\,${\rm s}^{-1}\,$}}
\newcommand{\mpa}{\mbox{mag\,${\rm airmass}^{-1}\,$}}
\newcommand{\mpas}{\mbox{mag\,${\rm airmass}^{-2}\,$}}

\newcommand\micron{\mbox{$\mu$m}}%

\newcommand\arcdeg{\mbox{$^\circ$}}%
\newcommand\arcmin{\mbox{$^\prime$}}%
\newcommand\arcsec{\mbox{$^{\prime\prime}$}}%

\newcommand\farcmin{\mbox{$.\mkern-4mu^\prime$}}%
\newcommand\farcsec{\mbox{$.\!\!^{\prime\prime}$}}%

\title{New optical telescope projects at Devasthal Observatory} 

\author{Ram Sagar, Brijesh Kumar, Amitesh Omar and A. K. Pandey
\skiplinehalf
 Aryabhatta Research Institute of Observational Sciences,
 Manora Peak, Nainital 263 129, India
}

\authorinfo{For correspondence (E-mail: brij@aries.res.in; Telephone: +91 5942 35136)}
\begin{document} 
  \maketitle 

%%%%%%%%%%%%%%%%%%%%%%%%%%%%%%%%%%%%%%%%%%%%%%%%%%%%%%%%%%%%% 
\begin{abstract}
Devasthal, located in the Kumaun region of Himalayas is emerging as one of the best optical 
astronomy site in the continent. The minimum recorded ground level atmospheric seeing at the site
is 0\farcsec6 with median value at 1\farcsec1. Currently, a 1.3-m fast (f/4) wide field-of-view
(66\arcmin) optical telescope is operating at the site. In near future, a 4-m liquid mirror 
telescope in collaboration with Belgium and Canada, and a 3.6-m optical telescope in collaboration
with Belgium are expected to be installed in 2013. The telescopes will be operated by
Aryabhatta Research Institute of Observational Sciences. The first instruments on the 3.6-m
telescope will be in-house designed and assembled faint object spectrograph and camera. The
second generation instruments will be including a large field-of-view optical imager, high 
resolution optical spectrograph, integral field unit and an optical near-infrared spectrograph. 
The 1.3-m telescope is primarily
used for wide field photometry imaging while the liquid mirror telescope will see a time bound
operation to image half a degree wide strip in the galactic plane. There will be an aluminizing
plant at the site to coat mirrors of sizes up to 3.7 m. The Devasthal Observatory and its
geographical importance in between major astronomical observatories makes it important for
time critical observations requiring continuous monitoring of variable and transient objects
from ground based observatories. The site characteristics, its expansions plans and first
results from the existing telescope are presented.
\end{abstract}

%>>>> Include a list of keywords after the abstract 

\keywords{Astronomical site, Optical telescope and instrumentation, Devasthal Observatory, 
          Atmospheric seeing, Sky darkness, Liquid mirror telescope}

% sec:intro
%_____________________________________________________________________________

\section{introduction} \label{sec:intro}  % \label{} allows reference to this section

\begin{figure}
\centering
\includegraphics[angle=90,width=12cm]{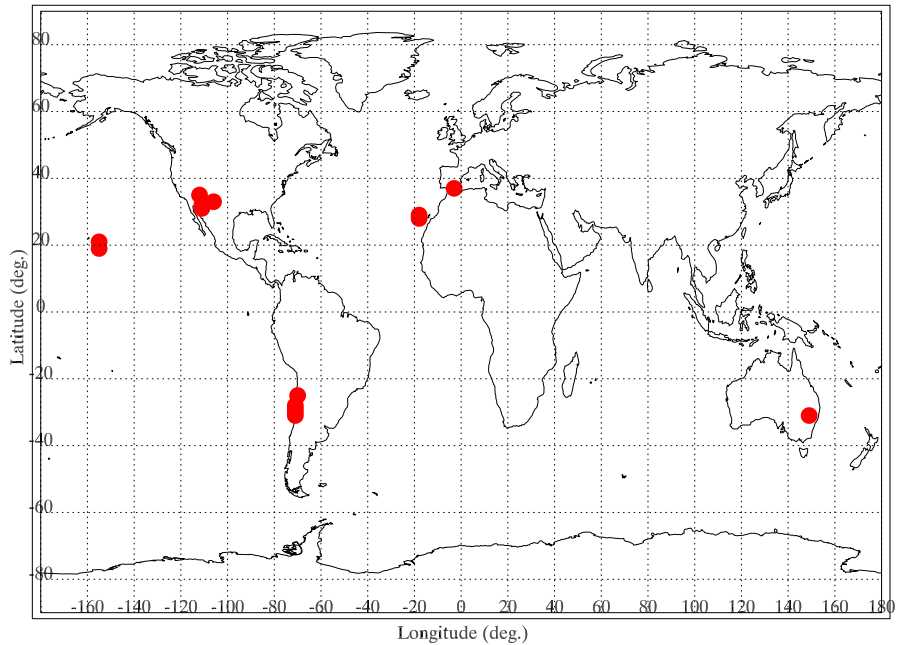}%
\caption{Geographical location of the existing 4-m class optical telescopes located
         at Mauna Kea ($-$155\arcdeg), Kitt Peak ($-$111\arcdeg), La Silla ($-$71\arcdeg),
         La Palma ($-$18\arcdeg), Calar Alto ($-$3\arcdeg) and Siding Spring($+$149\arcdeg) 
         are marked 
         with red circles. The Devasthal ($+$79\arcmin) lies at the middle of 12 hours gap in 
         geographical longitudes of Calar Alto in the West and Siding Spring in the East.}
\label{fig:map}
\end{figure}

Aryabhatta Research Institute of Observational 
Sciences (acronym ARIES)\cite{2004FL.21..30P,2006BASI...34...37S}\, an
autonomous research institute under the Department of Science and Technology, 
Government of India, has taken initiative to establish moderate size 
(up to 4-m class) optical telescopes at Devasthal in Nainital, India. 
The Devasthal\footnote[1]{Meaning abode of God, Longitude : 79\arcdeg41\arcmin04\arcsec E, 
Latitude : 29\arcdeg21\arcmin40\arcsec N, Altitude : 2450 m} is located in the foothills of 
central Himalayas. 
The technological advancements and the availability of sensitive detectors have made 
moderate size optical telescopes extremely valuable even today due to the increased 
level of performance, minimal maintenance and specific scientific goals. Furthermore, 
such telescopes at a good 
astronomical site have several advantages over very large (10-m class) and giant (30-m class) 
ones, e.g. in efficiency, availability, survey work, serendipitous discovery and time-critical
observations\cite{2000CS..78.1076P,1998BASI...26..417G}\,. A wide field 1.3-m optical 
telescope has successfully been installed at Devasthal in the year 2010 and another 3.6-m optical 
telescope having active optics technology is being built. The Devasthal will also host
a 4-m optical telescope having liquid mirror technology, being constructed in collaboration 
with Belgium and Canada. As these telescopes are being installed at a good observing site and 
also are being located at a crucial geographical longitude on the globe, 
see Figure~\ref{fig:map}, the Devasthal will 
have an added advantage for a number of time-critical observations of cosmic events. The
Devasthal Observatory is expected to significantly increase the access of moderate size optical 
telescopes to the Indian astronomical community. The need for an easy access to a well 
instrumented moderate sized optical telescopes is further underscored by the expected many-fold 
jump in the required optical observing time by Indian Astronomers in future on account 
of the upcoming first Indian Multi-wavelength Astronomical 
Satellite (ASTROSAT)\footnote[2]{http://www.iucaa.ernet.in/astrosat/} and already operational 
Giant Meterwave Radio Telescope (GMRT)\footnote[3]{http://gmrt.ncra.tifr.res.in/} at Pune, India. 

The next section describes the Devasthal as an astronomical site while the
subsequent sections provide scientific objectives, technical descriptions
and the preliminary results on the 1.3-m, 3.6-m and 4-m optical telescope projects. 
%todo : participation of and usage by national and international communities can be
%highlighted 

% fig:map
%__________________________________________________________________________

% sec:site
%_____________________________________________________________________________

\section{devasthal site} \label{sec:site}  % \label{} allows reference to this section

\begin{figure}
\centering
\includegraphics[width=6cm]{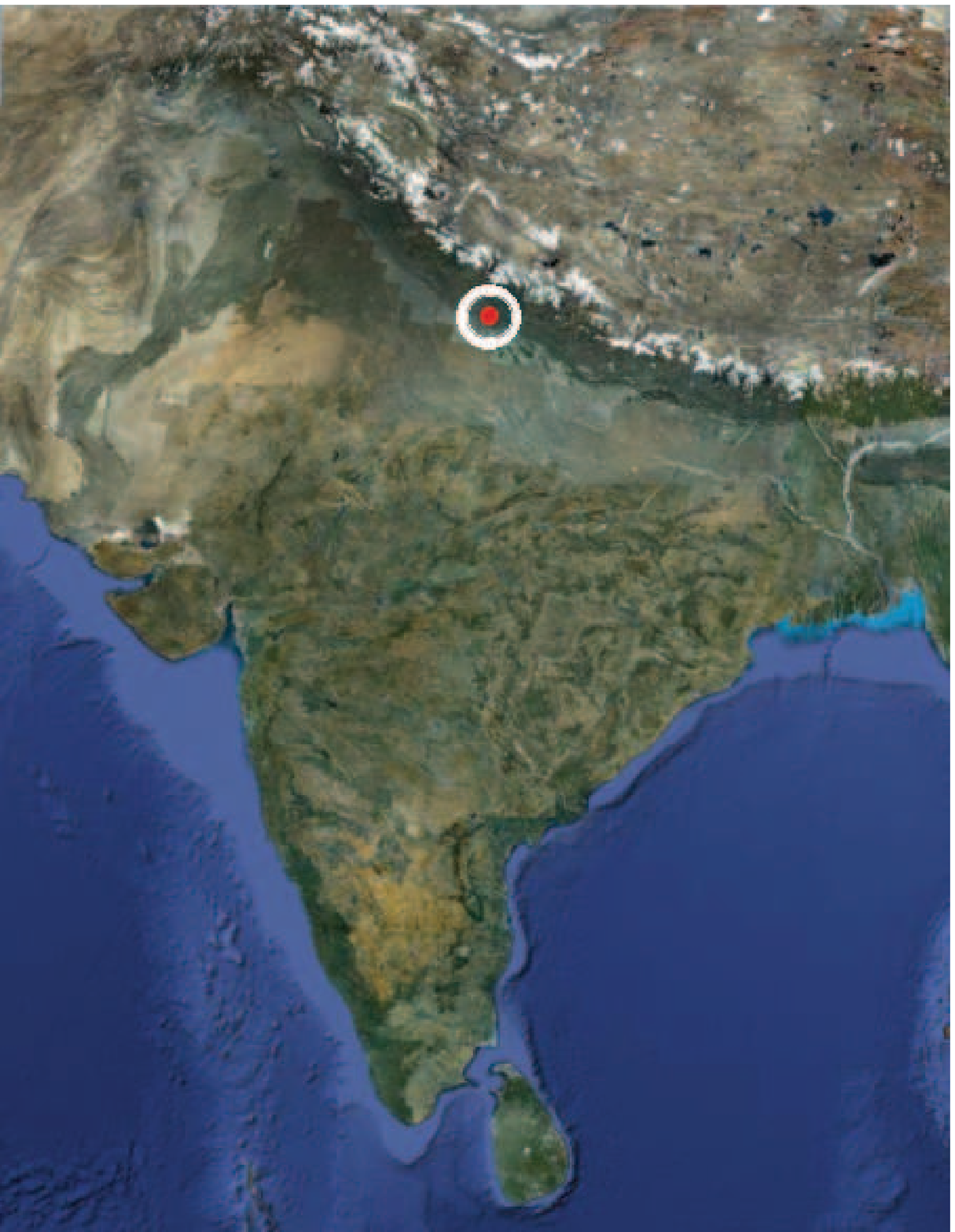}%
\hspace{10pt}\includegraphics[width=9.8cm]{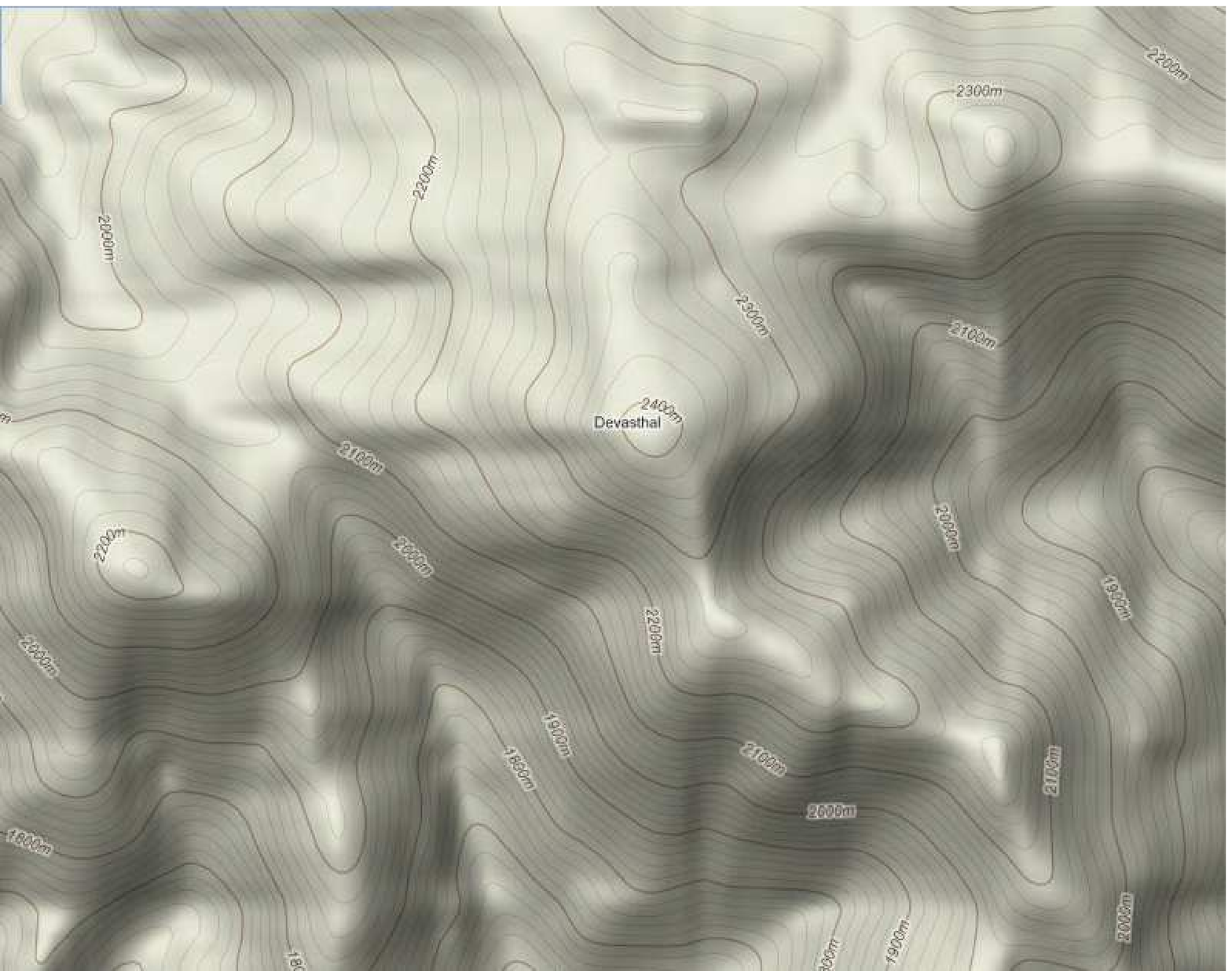}%
\caption{Geographical location of the astronomical site - Devasthal in the Indian sub-continent
         (left). The location is marked with a red dot encircled in white color. Contour map of 
         the Devasthal site (right). The area is about 4 km on a side and the Devasthal Peak lies 
         in the center. Courtesy Google Map.}
\label{fig:site}
\end{figure}

An extensive site survey conducted in the Kumaun region of the central
Himalayas identified Devasthal as a potential site for astronomical observations at
optical and near infrared wavebands. It is well connected by road with the 
Manora Peak\footnote[4]{Longitude : 79\arcdeg27\arcmin26\arcsec E, 
Latitude : 29\arcdeg21\arcmin39\arcsec N, Altitude : 1927 m} located in the vicinity of 
Nainital\cite{1972oams.conf...20S,1999CS..77..643P}\,. The Devasthal and Manora Peak have a 
longitudinal separation of about 18 km and both the sites are located in the Nainital District 
of Uttarakhand state in India, and they are operated by ARIES. The geographical location 
of these astronomical sites are shown in Figure~\ref{fig:site}. 

The site survey work in the Kumaun region was initiated in 1980 and the meteorological 
observations 
at a number of places were carried out during 1982-91\cite{2000BASI...28..429S}\,. The Devasthal
peak was located far from any urban development and there was no mountain having higher peak 
within an aerial distance of 1 km. The Devasthal also offered about 210 useful spectroscopic 
nights, out of which a good fractions (80\%) of the nights are of photometric quality. A detailed 
characterization campaign of Devasthal was performed during 1998-1999\cite{2000A&AS..144..349S}\,. 
The air temperature at Devasthal varies between $-5$ to $+22$\arcdeg\,C over the year while the 
intranight temperature variation is less than 2\arcdeg\,C. The relative humidity is below 60\%
during spectroscopic nights, while it can go to very high values during rainy months of July, 
August and September. The wind speed at Devasthal is below 5 \mps for about 85\% of the time. 

The atmospheric extinction coefficients measured 
during 1998-91\cite{1999BASI...27..601M} with a solid state stellar photometer mounted on 
a 52-cm reflector resulted in a mean values of $0.49\pm0.09$, $0.32\pm0.06$, $0.21\pm0.05$, 
$0.13\pm0.04$, and $0.08\pm0.04$ \mpa in the Johnson $UBVR$ and $I$ band respectively.
The minimum recorded value of extinction coefficients are 0.38, 0.22, 0.12 and 0.06 \mpa 
respectively in $UBV$ and $R$ bands. The atmospheric extinction at Devasthal measured 
recently in December 2010, using CCD camera mounted with a 1.3-m optical telescope, gives
a values of 0.24, 0.14 and 0.08 \mpa respectively in Johnson $BV$ and Cousins $R$
band; while the sky brightness is measured to be 21.2 \mpas in $V$ 
band\cite{2011CS.101.1020P}\,.  

The seeing measurements carried out on 80 nights during 1998-1999 with a Differential Image 
Motion Monitor (DIMM) using a 38-cm telescope with the mirror about 2 m above the ground, 
yielded a median seeing estimate of about 1\farcsec1; the 10 percentile values lie between 
0\farcsec7 to 0\farcsec8 (mean = 0\farcsec75) while for 35\% of the time the seeing was better 
than 1\arcsec \cite{2000A&AS..144..349S,2001BASI...29...39S}\,. 
The inference of a true median seeing for a site depends on many other factors, for example, 
statistics, instrument used and data reduction procedures. In the present case, the seeing 
measurements are made using DIMM with a 10 ms exposure time, using a hole separation of 24 cm 
and a pixel separation of 60. It is observed that a finite, say 10 ms exposure is likely to 
under estimate the seeing by 10 percent. This shall degrade our median seeing estimate to 
about 1\farcsec2 Full Width at Half Maximum (FWHM). Furthermore, the above 
median seeing (1\farcsec1) estimate are performed at 2 m above ground level and it is observed 
that a seeing of 0\farcsec86 \, is contributed by the 6 to 12 m slab of the Devasthal Earth's 
atmosphere above the ground\cite{1999A&AS..136...19P,2001BASI...29...39S}\,. Consequently, 
if a telescope is placed at around 8 meters 
above ground, a median seeing of sub-arcsec (about 0\farcsec6) could be expected at Devasthal. 
In addition, the free air seeing gets further degraded by the local air turbulence due to dome 
design. This effect may be from a few percent to tens of percent. It, is therefore suggested 
that for a height of 8 m above ground the best seeing (10 percentile) value for Devasthal 
can be taken as 0\farcsec7 FWHM. 

%fig:site
%__________________________________________________________________________

The infrastructure development has been carried out extensively at Devasthal site. The nearest 
state road is at 3.5 km away from the Devasthal Peak. The institute has built a 6 m wide metalled
road up to the peak. There is a high speed 2.4 GHz microwave link of 18 Mbps bandwidth between
Manora Peak and Devasthal. The requirements of water is met by deep borewell and through 
rainwater recharging pits.

%fig:dfot
%__________________________________________________________________________
\begin{figure}
\centering
\includegraphics[width=6cm]{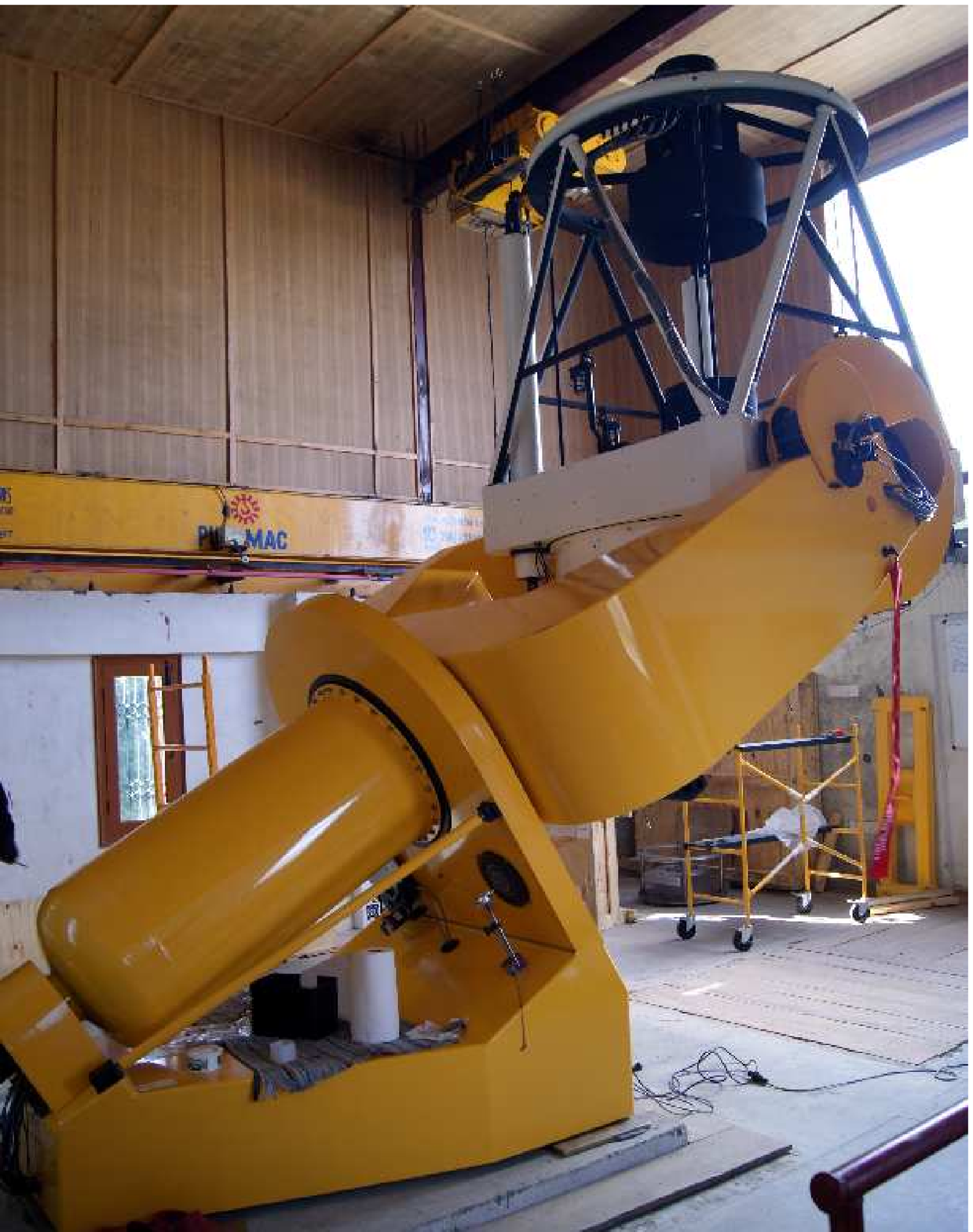}%
\hspace{10pt} \includegraphics[width=10.1cm]{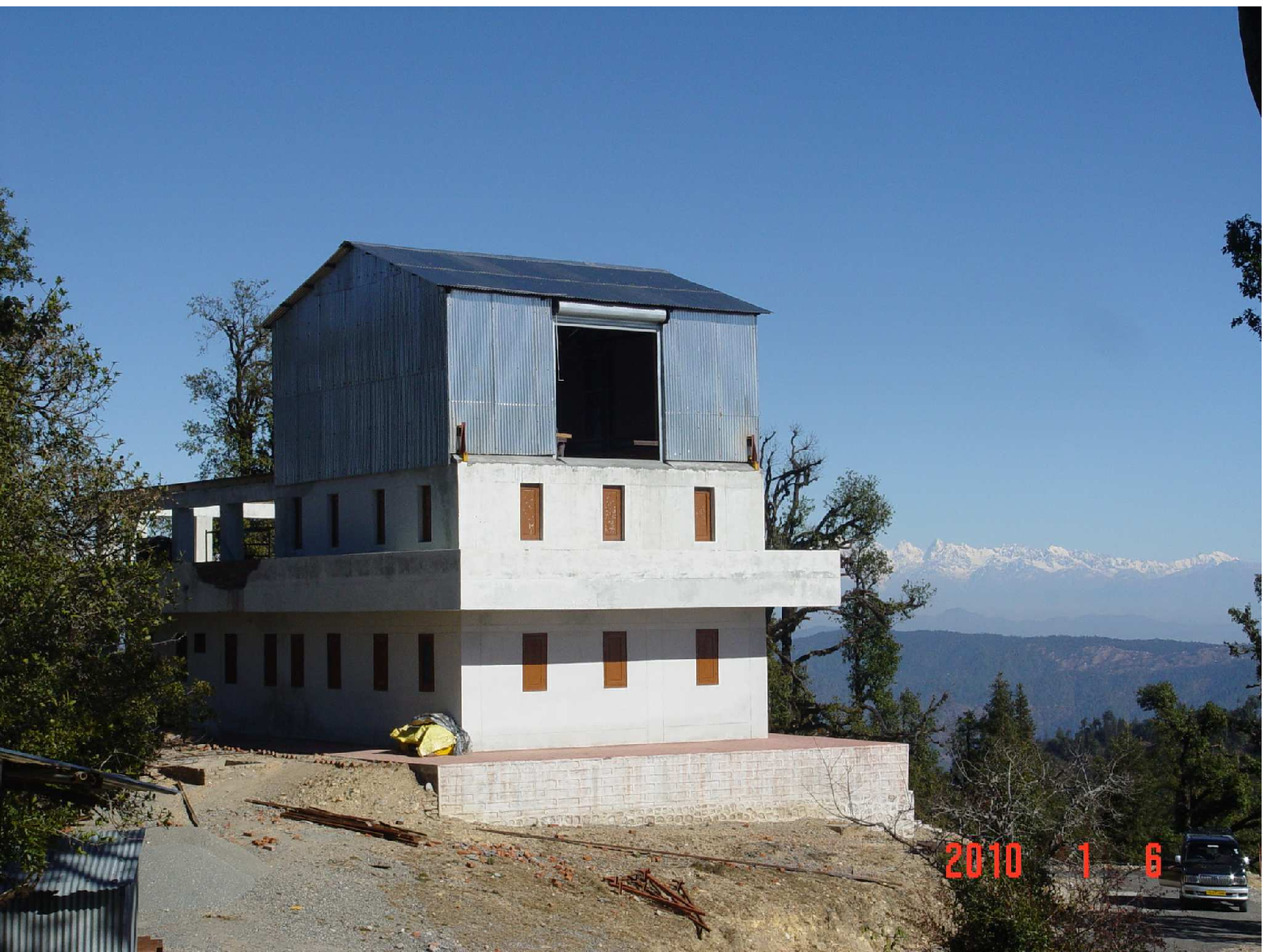}%
\caption{The 1.3-m Devasthal Fast Optical Telescope at Devasthal (left) and its roll-off roof 
         enclosure (right).}
\label{fig:dfot}
\end{figure}

% sec : 130cm-DFM
%___________________________________________________________________________

\section{1.3m DFOT - Devasthal Fast Optical Telescope} \label{sec:opt}

In October 2010, a new modern 1.3-m DFOT has been
installed successfully at Devasthal\cite{2011CS.101.1020P}\,. A picture of the
telescope and its enclosure is given in Figure~\ref{fig:dfot}. In order to avoid degradation 
of seeing due to local environments, the telescope is mounted 3 m above the ground and the 
enclosure has a roll-off roof design. The telescope design of the 2-mirror
Ritchey-Chr{\'e}tien optics along with a single element corrector is optimized to deliver 
a fast beam (f/4, plate scale of 40\arcsec\,mm$^{-1}$) and a naturally flat-field of 
66\arcmin\,diameter at the axial Cassegrain focus\cite{2000SPIE.4003..456M}\,. It is therefore
suitable for wide-area survey of a large number of point as well as extended sources.
The telescope optics can deliver images with 80\% encircled-energy (E80) diameter of 0\farcsec6 at 
visible wavebands (or equivalently a Gaussian Point Spread Function (PSF) of 0\farcsec4 FWHM).  
Without autoguider the tracking accuracy of the telescope is better than 0\farcsec5
in an exposure of 300 s up to a zenith distance of 40\arcdeg. The pointing accuracy
of the telescope is better than 10\arcsec\,Root Mean Squared (RMS) for any point in the sky. 
Further technical details on the as-designed specifications of the telescope system are given
elsewhere\cite{2010ASInC...1..203S}\,. The main scientific
objective is to monitor optical and near infrared (350-2500 nm) flux variability
in the astronomical sources such as transient events (Gamma-ray bursts, supernovae),
episodic events (active galactic nuclei, X-ray binaries and cataclysmic variables), 
stellar variables (pulsating, eclipsing and irregular), transiting extrasolar planets - and to 
carry out photometric and imaging surveys of extended astronomical sources, e.g. HII regions,
star clusters, and galaxies. Further details on the scientific objectives can be
found elsewhere \cite{2006BASI...34...37S}.

During commissioning phase, the telescope was equipped with a 13.5 \micron\, pixel, 2k$\times$2k 
Andor iKon Camera\footnote[5]{http://www.andor.com (Model DZ436)} which covers a square area of 
about 18\arcmin\, sky on a side. A set of Johnson-Cousins $B$, $V$, $R$ and $H_{\alpha}$ filters, 
circular in size and providing unvignetted field of 18\arcmin\, diameter on CCD were available.
For near-zenith observations, in about 2\arcsec\, FWHM PSF, we could reach a signal-to-noise 
ratio of 5 for 20 mag in $B$ and 21 mag in $V$ and $R$ in five summed images of 10s each. 
The atmospheric extinctions in $B$, $V$, and $R$ are estimated as $0.16\pm0.01$, $0.11\pm0.01$ 
and $0.05\pm0.02$ \mpa respectively on 30th November 2010. 

\begin{figure}
\centering
\includegraphics[width=9cm]{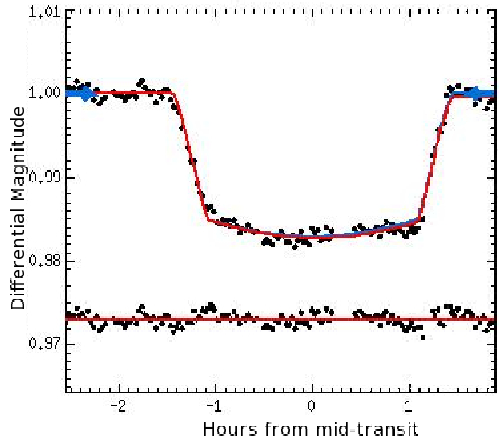}%
\caption{The R-band differential light curve of the transiting
         extrasolar planet WASP12 ($V \sim$ 11.7 mag star) observed on 5$^{\rm th}$ February 2011
         with the 1.3-m DFOT at Devasthal, Nainital. The model is overplotted and a typical error 
         in each data point is of the order of 1 mmag.}
\label{fig:wasp}
\end{figure}

In order to know the detection limits of low-amplitude flux variations in
brighter ($\sim 10$ mag) celestial sources set by scintillation in the Earth's atmosphere, we
also carried out photometric observations of a known transiting extrasolar planet
WASP-12 ($\alpha_{\rm J2000} = 06^{\rm h} 30^{\rm m} 32^{\rm s}$,
$\delta_{\rm J2000}=$ 29\arcdeg 40\arcmin 40\arcsec, $V$ = 11.7 mag).
We used 16\micron\, pixel, 512$\times$512 Andor iXon Camera\footnote[6]{EMCCD Model DU897} which 
covers a square area of 7\arcmin\, on a side. On 5$^{\rm th}$ February 2011, we recorded 
a set of 3300 CCD frames
of 5 s each in Cousins $R$-band during a continuous observations for 4.5 hours. The
observations were made without auto guider. The data reduction procedures are described 
elsewhere\cite{2009MNRAS.392.1532J} and the differential light curve was generated 
using ensemble photometry by employing four comparison stars. The differential light curve had
a typical photometric accuracy of 3 mmag. To improve the signal-to-noise ratio, we co-added 20
frames of 5 s each and this co-added differential light curve of the WASP-12 transiting system 
along with the model fit indicates a photometric precision of 1 mmag for a 11.7 mag star (see 
Figure~\ref{fig:wasp}). 
As a comparison, a similar observations using 104-cm Sampurnanand Telescope at Manora Peak, we get
an accuracy of about 3 to 4 mmag. Hence the 1.3-m DFOT at Devasthal would be suitable
for the scintillation limited science programs requiring a detection of few mmag
on a time scale of hrs (e.g. exoplanet search and AGN variability).

A detailed report on commissioning of the 1.3-m DFOT can be found 
elsewhere\cite{2012ASInC...3..203S}\,.

%fig:wasp
%__________________________________________________________________________

% sec : 360cm-AMOS
%_____________________________________________________________________________

\section{The 3.6m DOT - Devasthal Optical Telescope}

The 3.6m DOT facility consist of a modern 3.6 meter optical telescope, 
a suite of instruments, an observatory with an aluminising plant, a control room and a data center. 
The telescope will have a number of instruments providing high resolution spectral and imaging 
capabilities at visible and near-infrared bands. In addition to optical studies of a wide 
variety of astronomical topics, it will be used for follow-up studies of sources identified 
in the radio region by GMRT and UV/X-ray by ASTROSAT. The contract to design, manufacture, 
integration, testing, supply and installation at 
Devasthal of the 3.6m DOT is awarded to the Advanced Mechanical and Optical System (AMOS)
Belgium\footnote[7]{http://www.amos.be} in March 2007 and as on May 2012, the telescope is 
fully assembled with real mirrors at the AMOS workshop and the tests to verify the telescope 
performance are being performed. 

\subsection{Telescope}
\begin{figure}
\centering
\includegraphics[width=12cm]{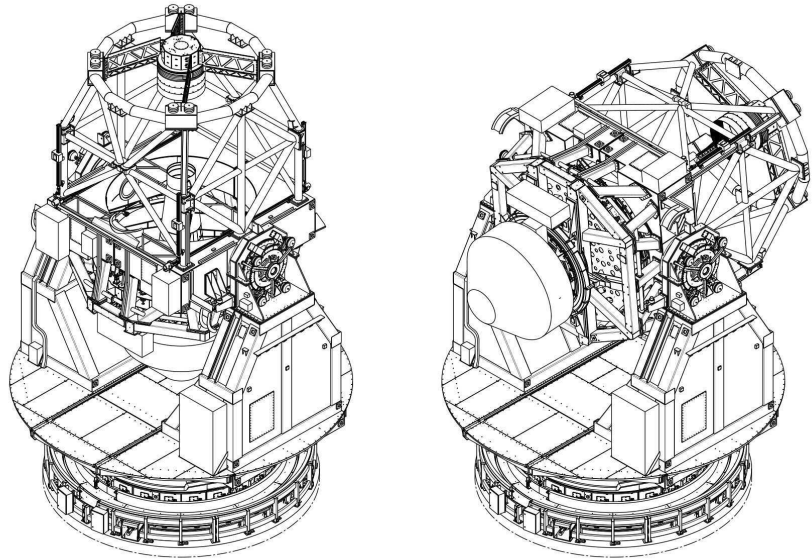}%
\caption{The isometric view of the final design of the 3.6-m DOT - zenith position (left), horizon
         position at 85 deg zenith distance (right). The telescope is 13.3 m high and it weighs
         149.3 ton.}
\label{fig:dot}
\end{figure}

A computer rendering of the complete as-built telescope is shown in Figure~\ref{fig:dot}. The 
telescope is a two-mirror RC system with f/9 configuration and 
an alt-azimuth mount\footnote[8]{More details on the as-designed technical 
specifications as well as a general description of the telescope can be found 
elsewhere\cite{2007NASL.30..209P,2008SPIE.7012E...8F,2010ASInC...1..203S,2012SPIE.8444..67F}}. The 
primary mirror is made from a thin meniscus ZERODUR\footnote[9]{http://www.schott.com} glass 
having a focal ratio of f/2, a clear 
aperture of 3.6 m diameter and a thickness of 165 mm. The secondary mirror is made from a 
plano-plano Astrositall glass having a diameter of 0.98 m. Both, the primary and secondary 
mirrors of the telescope are polished at the Lytkarina Optical Glass 
Factory (LZOS)\footnote[1]{http://www.lzos.ru}, Russia and the polishing accuracy (RMS wavefront 
error at 600 nm) of 35 nm for primary and 30 nm for secondary has been 
achieved\cite{2012SPIE.8450..176F}\,. In order to exploit the 
best seeing (0\farcsec7) at Devasthal, the optics of the telescope is designed to deliver 
E80 in less than 0\farcsec45 diameter (or equivalently a WFE RMS $<$ 210
nm) in a 10\arcmin\, arcmin Field of View (FoV) over 350 nm to 1500 nm wavelength range without 
corrector. The RMS WFE of as-built telescope at worst altitudes is computed as 137 nm for side 
port and 184 nm for axial port. A preliminary speckle imaging tests performed at the AMOS 
workshop suggest that the as-built optics can deliver images with E80 better than 0\farcsec3 
diameter\cite{2012SPIE.8444..102F}\,. 

%fig:dot
%__________________________________________________________________________

%fig:ins
%__________________________________________________________________________
\begin{figure}
\centering
\includegraphics[width=6cm]{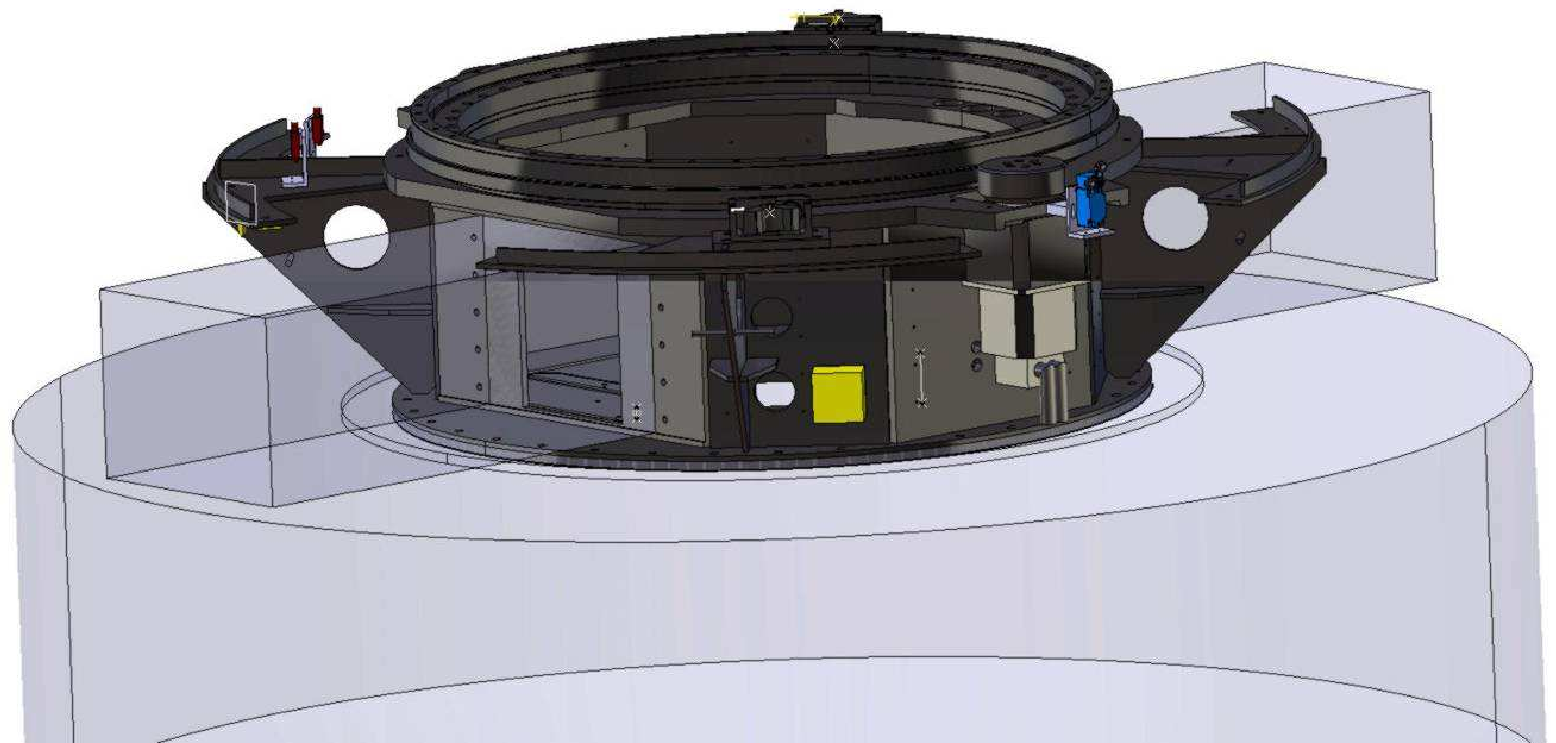}%
\includegraphics[width=6cm]{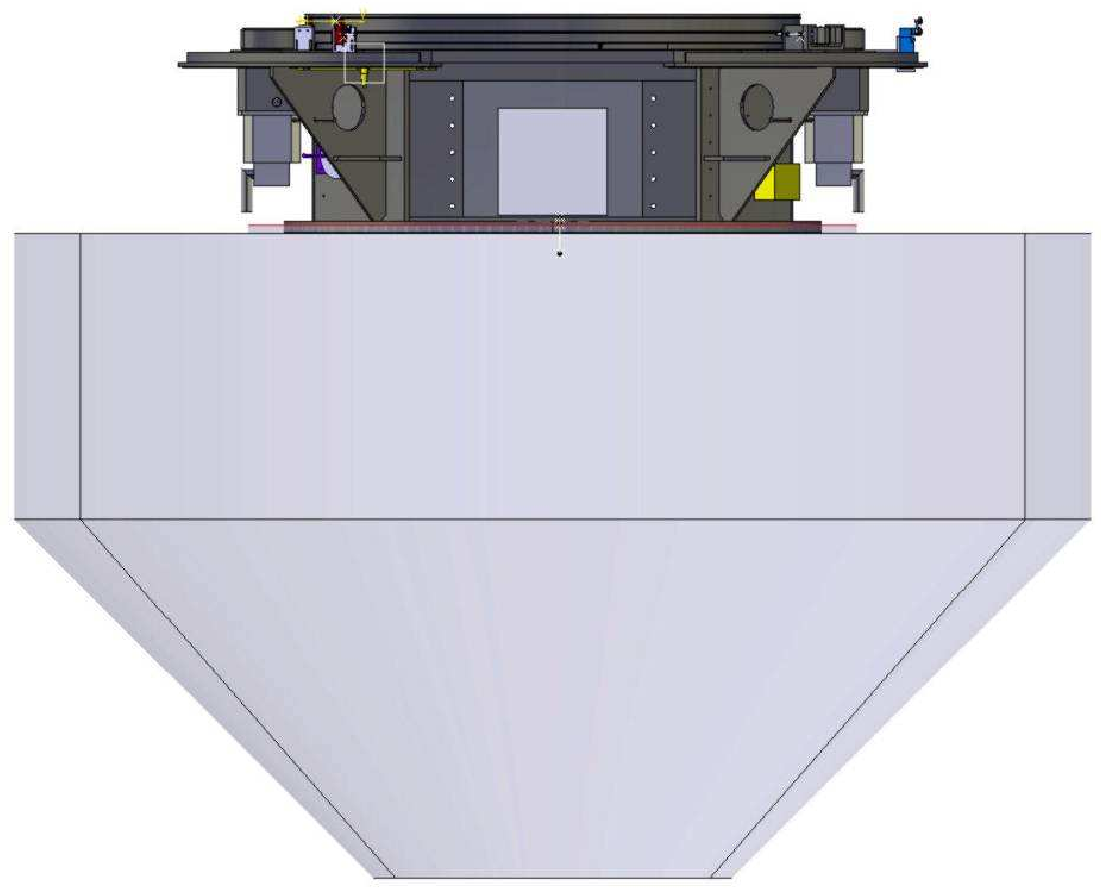}%
\caption{The instrument mounting interface of 3.6-m DOT is shown. The computer rendering of 
         instrument envelope for two side ports and one axial port is also shown.}
\label{fig:ins}
\end{figure}

%fig:enc1
%__________________________________________________________________________
\begin{figure}
\centering
\includegraphics[width=10cm]{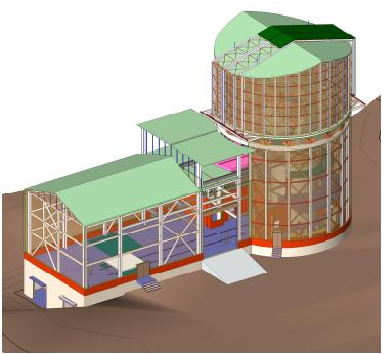}%
\caption{The concept design of the 3.6-m DOT building.}
%\vskip 8cm
\label{fig:enc1}
\end{figure}

The azimuth, altitude and other bearings of the telescope uses latest technology available
in the market\cite{2012SPIE.8444..150F}\,. The tracking performance of the telescope is better 
than 0\farcsec1 RMS for one minute in open loop for wind inside the dome of less 
than 3 \mps\, and in close loop it is 0\farcsec11 for less than one hour. In open loop and for wind 
inside the dome of 5 \mps\, the tracking accuracy is $\sim$0\farcsec5 peak in 15 minute. 
The preliminary test results performed at AMOS workshop obtains a value of 0\farcsec2 
\, RMS in open loop for 10 minutes. The pointing accuracy is better than 2\arcsec\, RMS 
for any point in the sky with elevation greater than 10\arcdeg\, and wind inside the dome of less 
than 5 m/s. For an offset of 10\arcmin\, the pointing accuracy is 0\farcsec1 RMS with wind inside 
the dome of less than 3 \mps\,. The telescope can slew in azimuth with 2\dps\, and in altitude 
with 1\dps\,.

The telescope has an auto-guiding and wavefront sensor unit which are used to calibrate 
the telescope's performance with the help of a bright star picked up from the annular 
(31\arcmin-35\arcmin\,diameter) region of field of view. The telescope is also equipped with 
the active optics system (AOS)\cite{2012SPIE.8444..186F} which is a low frequency system 
that detects and corrects deformations, aberrations or any other phenomenon that degrade 
the image quality of the telescope. The AOS consists of – a wave front sensor, 69 pneumatic 
active M1 support systems, M2 hexapod and the telescope control 
system\cite{2012SPIE.8451..82F} that acts as interface between each element.

The telescope is provided with instruments mounting cube with one axial port with 30\arcmin\, FoV 
and two side ports with 10\arcmin\, FoV, see Figure~\ref{fig:ins}. 
The axial port can also be fitted with a detachable 30\arcmin\, wide 
field three-lens corrector unit. The main axial port is designed for instruments weighing 2000 kg, 
while side ports are designed for mounting instruments with 500 kg each. The Cassegrain end can 
also take imbalance 
for 2000 Nm on altitude axis and 400 Nm on rotator axis. The axial port instrument envelope is a 
cylindrical cum conical space of 1.8 m height below the axial instrument flange. A cylindrical 
diameter of 3 m from flange to 0.8 m and a decreasing conical diameter from 3 m 
to 1 m corresponding to the distance below the flange from 0.8 m to 1.8 m. The side port instrument 
envelope is a cuboid of sides 380 mm, 500 mm and 1000 mm.

% subsection : 
%_______________________________________________________________________________

\subsection{Telescope enclosure and extension building}
The contract to design, engineering, procurement assistance, inspection, and testing of 
enclosure and auxiliary building for the telescope is awarded to Precision Precast Solutions 
Private (PPS) limited, Pune\footnote[2]{http://ppspl.com}. The contract for 'manufacture, supply, 
erection and commissioning of telescope enclosure structure and equipments is awarded 
to M/s Pedvak Cranes Privated Ltd., Hyderabad. The final design for the telescope 
house is shown in Figure~\ref{fig:enc1}. The telescope building structure is divided into 
three parts - the rotating dome (fully insulated steel framed structure), the dome building 
and the extension building. The design of the civil work up to plinth level of the telescope 
site is done by PPS, while the civil work up to plinth level was completed by local builders. 

\begin{figure}
\centering
\includegraphics[width=16cm]{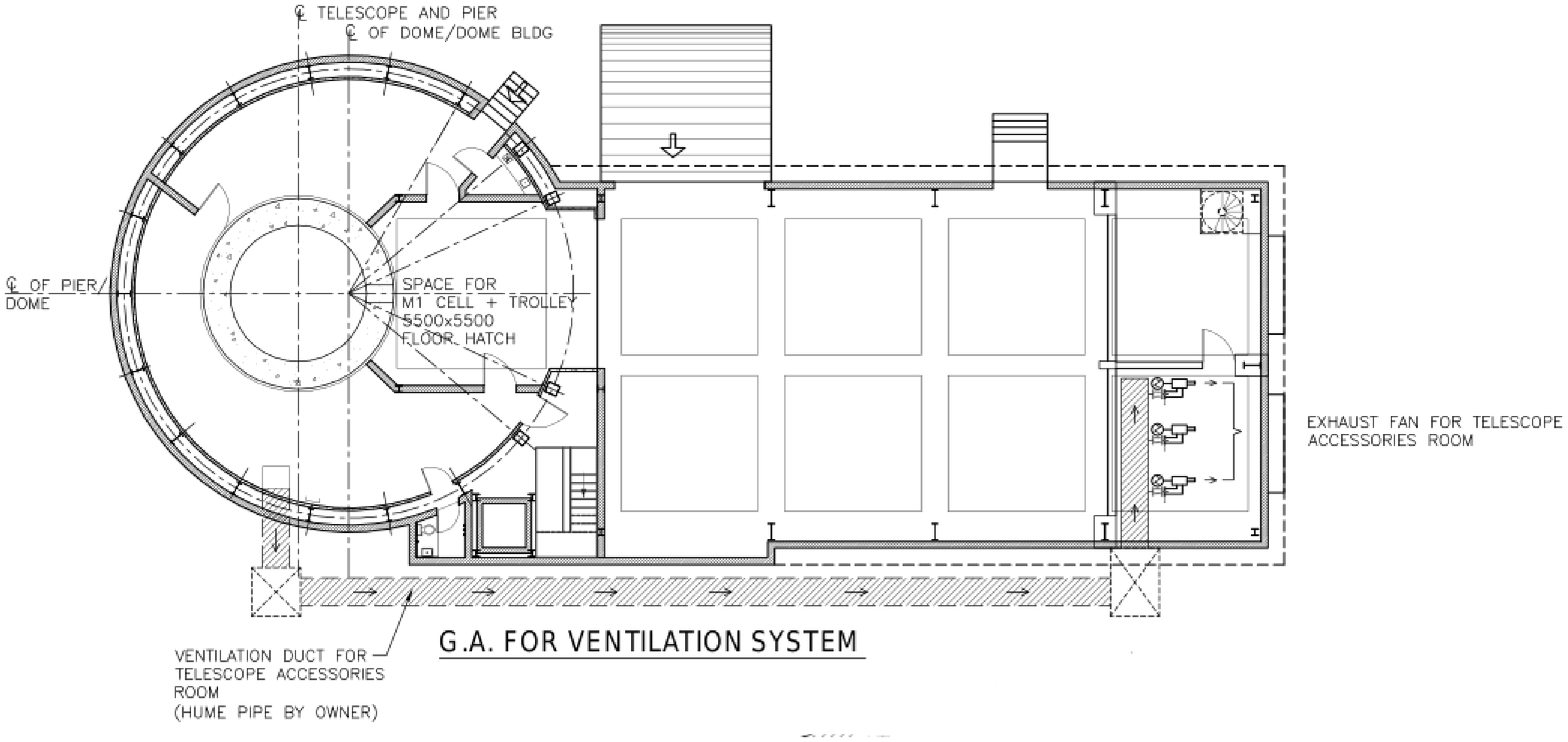}%
%\vskip 8cm
\caption{The floor plan of the 3.6-m DOT dome and the extension building. The extension 
         building will house the aluminising plant (see \S\ref{sec:coat}.)}
\label{fig:enc2}
\end{figure}

Considering space limitation at the site, a 16.5 m diameter dome building and an 
off-centered telescope pier has been envisaged, see Figure~\ref{fig:enc2}. The gyration radius 
of the telescope is 5.749 m. The foundation and structure of the telescope pier (cylindrical, 
8.2 m high above plinth level and 7 m width) is fully isolated from the dome structure to avoid 
transfer of vibrations. The level of primary mirror will be at about 14.2 m above the plinth level
of the ground. The as-designed analysis indicates that the natural frequency of the bare pier 
is 25.44 Hz for first lateral mode in Z-direction, while for both the pier and the telescope 
is 14.259 Hz. The first Eigen frequency of the telescope is 7.4 Hz as estimated from
finite element analysis. To avoid degradation of seeing due to local heating, two separate 
ventilation ducts - one from telescope floor and another from telescope technical room along 
with the exhaust fans have been provided (see Figure~\ref{fig:enc2}). Further details on the 
enclosure design of the telescope can be found elsewhere\cite{2012SPIE.8444..152F}\,.  

An observatory control and data archive system for the telescope is being developed in-house
at ARIES. The synchronization of telescope motions with the dome motions is a bit tricky in 
case of 3.6-m DOT as the telescope center and the dome center are separated by about 1.85 m. The
study found that there would be two regions of avoidance near zenith (blind spot), one for the 
telescope and another for the dome.  

%fig:enc2
%__________________________________________________________________________

% subsection : 
%_______________________________________________________________________________

\subsection{Aluminising plant} \label{sec:coat}
The first aluminization of the primary mirror shall be done at Devasthal before final performance 
test of the telescope. A contract for design, manufacture and commissioning of an aluminium 
coating plant for mirrors up to 3.7m diameter has been awarded to Hind High Vacuum (HHV) Company 
Private Limited, Bangalore\footnote[3]{http://www.hhv.in}. The aluminium sputtering plant has been 
designed, manufactured, assembled and tested at the HHV workshop. The tests performed at the 
workshop indicates that the aluminium deposition of 1000\AA\, can be done with uniformity 
of 25\AA. The reflectivity measurement done on sample mirrors have resulted in a value 
of $92\pm2$\% for the wavelength range 350-850 nm. The coating plant has been shipped to the 
Devasthal site and it will be installed in the extension building of the telescope house. 
Further details on the coating plant can be found elsewhere\cite{2012SPIE.8444..80F}\,.  

% subsection : 
%_______________________________________________________________________________

\subsection{Instrumentation}

The first generations focal plane instruments are a faint object spectrograph
and camera, a high resolution fiber-fed optical spectrograph, an optical--near infrared
spectrograph and imager, and a CCD optical imager (see Table~\ref{tab:ins}).

The Faint Object Spectrograph and Camera (FOSC) is a focal reducer instrument.
The instrument shall work in imaging and spectroscopic mode. The instrument will have
imaging capabilities with one pixel resolution of less than 0\farcsec2 in the
FoV of $\sim$ 14\arcmin\, $\times$ 14\arcmin\, of the telescope, and
low-medium spectroscopy with spectral resolution (250-4000)
covering the wavelength range from 350 nm to 900 nm. A computer simulation indicate
that we can image a 25 mag star in $V$ band with an hour of exposure time.
The optical and mechanical design of the instrument has been completed in-house at
ARIES. Further technical details can be found elsewhere\cite{2012SPIE.8446..38F}\,.

\begin{figure}
\centering
\includegraphics[width=8cm]{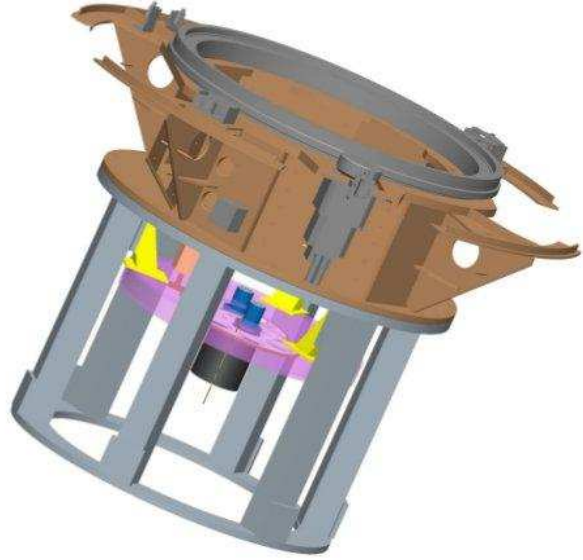}%
%\vskip 8cm
\caption{The mechanical design interface for the optical imager.}
\label{fig:oi}
\end{figure}

An optical imager with a 15 micron pixel, 4k$\times$4k back-illuminated CCD detector, liquid 
nitrogen cooling, full frame window mode operation, and the associated control electronics has 
also been proposed as a first light instrument. A contract for assembly and integration of the 
CCD camera has been awarded to Semiconductor Technology Associates, 
USA\footnote[4]{http://www.sta-inc.net/}. The mechanical interface for the camera is being
designed and manufactured in-house at ARIES. The mechanical interface for the camera was 
completed recently, see Figure~\ref{fig:oi}. This imager will primarily be used to verify the 
performance of telescope during the commissioning phase. The imager will cover a square 
area of 6\farcmin5$\times$6\farcmin5 on the sky. This instrument will have broadband 
Johnson-Cousins $UBVRI$ and $ugriz$ SDSS filters, as well as a few narrow-band filters. Once 
the other proposed instruments are ready for commissioning, the optical imager will 
be optimized for wide-field imaging at either 1.3-m DFOT or 3.6-m DOT.

%fig:oi
%__________________________________________________________________________

%  tab1
%____________________________________________________________________
  \begin{table}
  \small
  \caption{Technical specifications of the proposed Instruments for 3.6m DOT}
  \label{tab:ins}
  \centering

  \begin{tabular}{l l} \hline \hline
  \noalign{\smallskip}
   Parameters& value\\
  \noalign{\smallskip} \hline

  \noalign{\smallskip}
  \multicolumn{2}{l}{\bf Faint object spectrograph and camera:}\\
   Spectral coverage& 350-900 nm \\
   Field-of-view& 14\arcmin$\times$14\arcmin (imaging); 10\arcmin$\times$10\arcmin (spectroscopy)\\
   Image quality& 80\% energy in 0\farcsec4 diameter\\
   Resolving power& 250-2000 @1\arcsec\, slit-width with single grisms\\
   ~~~~~~~~~~~~~~~& 4000 @1\arcsec\, slit-width with VPH Gratings\\

   \\
   \multicolumn{2}{l}{\bf High-resolution optical spectrograph :}\\
   Spectral coverage& 380-900 mm\\
   Resolving power& 30k and 60k (fixed)\\
   Radial velocity stability& 20 \mps\, or better\\

   \\
   \multicolumn{2}{l}{\bf Optical and near-infrared spectrograph and imager:}\\
   Spectral coverage& 500 - 2500 nm\\
   Resolving power& 3000-4000\\
   Field of view & 7\arcmin\\
  \\
   \multicolumn{2}{l}{\bf Optical imager:}\\
   Spectral coverage& 300 - 900 nm\\
   Field of view & 6\farcmin5$\times$6\farcmin5\\
   Spatial resolution & 0\farcsec1 pix$^{-1}$
  \\
  \noalign{\smallskip}
  \hline
\end{tabular}
\end{table}

In order to complement the photometric and imaging capabilities with spectroscopy, a high 
resolution spectrograph  instrument is envisaged. The main science goals of the proposed 
spectrograph  are asteroseismology,  Doppler imaging of spotted stars, massive single
and binary stars and abundances studies. The instrument will be 
based upon a modern design using white pupil concept where a dual mode collimator forms white 
light image of the grating to be re-imaged by the camera with desired beam size. The concept 
will be similar to many contemporary high resolution spectrometers such as HERMES, HARPS, 
ESPaDoNs etc. The spectrograph will provide high-resolution spectra up to spectral resolution 
of 60k in single exposure and in the wavelength range 380-900 nm. The radial velocity stability 
is proposed to be better than 20 \mps. The instrument shall be capable of measuring spectrum with 
signal-to-noise ratio of 100 per 20 \kmps\, bin for an integration time of one hour for 
a star of $V \sim$ 16 mag.

A general purpose optical and near-infrared spectrograph and imager is proposed jointly by 
Tata Institute of Fundamental Research\footnote[5]{http://www.tifr.res.in}, Mumbai and ARIES for 
observations in the near-infrared bands between 500 nm to 2500 nm. It will use 
a 1024$\times$1024 Hawaii HgCdTe detector array manufactured by Rockwell International USA and will
have flexible optics and drive electronics that will permit a variety of
observing configurations. The primary aim of this instrument would be to obtain
broad and narrow band imaging of the fields as large as 6\arcmin$\times$6\arcmin\, and also
to use it as a long-slit spectrometer with moderate resolving
power ($\lambda/\Delta \lambda\sim 3500$) when attached to the telescope.
The proposed instrument when coupled with the 3.6-m telescope is expected to
reach the $5\sigma$ detection of 22.5 mag in $J$, 21.5 mag in $H$ and 21.0 mag
in $K$ with one hour integration.

%fig:ilmt
%__________________________________________________________________________
\begin{figure}
\centering
\includegraphics[width=8cm]{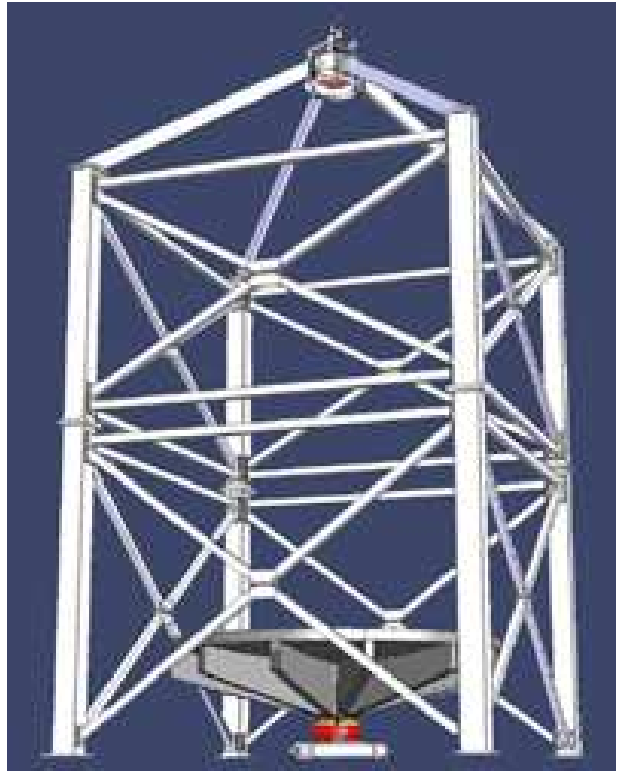}%
%\vskip 8cm
\caption{The as-designed view of the 4-m ILMT.}
\label{fig:ilmt}
\end{figure}

% sec : 4m-ILMT
%____________________________________________________________________________

\section{4-m ILMT - International Liquid Mirror Telescope}
The 4-m ILMT uses Liquid Mirror Technology and the mercury mirror of the ILMT 
will have a 4 m diameter with a focal ratio of f/2 (see Figure~\ref{fig:ilmt}). 
The ILMT is proposed to 
be installed at Devasthal as a joint collaboration between India, Belgium and Canada. 
It will perform as a transit telescope. A CCD detector shall be positioned at the prime 
focus of the telescope. The mirror being parabolic in shape needs a corrector to get a 
flat focal surface of about 30\arcmin\, diameter. The rotation of the Earth induces the motion
of the sky across the detector surface. The CCD detector works in a
Time delay integration mode, i.e. it tracks the stars by electronically
stepping the relevant charges at the same rate as the target moves
across the detector, allowing the integration as long as the target
remains inside the detector area. At the latitude of Devasthal,
a band of half a degree covers 156 square degrees, with 88 square degrees
being covered at high galactic latitude (b $>$ 30\arcdeg) including the
direction of the north galactic pole. The nightly
integration times are rather short, typically 120 s but it is possible
to co-add data from selected nights in order to get sky images of longer
integration times.

The expected limiting magnitudes are 24.5 at $U$, $B$ and $V$ bands, 23.5 at $R$ and $I$
bands and 22.3 at Gunn-z band. The expected database towards the Galactic Bulge
direction includes 10 million stars, 30000 variables, 8000 binaries,
8000 LPVs/SRVs, 5000 spotted RSCVn, 1400 RR Lyrae, 250 $\delta$-Scuti,
20 Cepheids, 50 yr$^{-1}$ microlenses and 5 yr$^{-1}$ Cataclysmic variables - providing
valuable inputs for the studies of stars, galaxies and cosmology.

Further details on 4-m ILMT can be found elsewhere\footnote[6]{http://www.aeos.ulg.ac.be/LMT/}.

% acknowledgements 
%_____________________________________________________________________________

\acknowledgments     %>>>> equivalent to \section*{ACKNOWLEDGMENTS}      
The work presented here is on behalf of a larger team associated with
the development of Devasthal Observatory. The authors are thankful to the Devasthal
Observatory staff for their assistance during observations taken with 1.3-m telescope. 
The overall guidance and direction of members of the project management board for the telescopes 
1.3-m DFOT and 3.6-m DOT is sincerely acknowledged.
The management board for 3.6m telescope consists of Prof. P. C. Agrawal (chairman), 
Prof. G. Srinivasan (vice-chairman), Prof. T.G.K. Murthy, Sri S. C. Tapde, 
Prof. S. Ananathkrishnan, Prof. S. N. Tandon, Prof. T. P. Prabhu, Prof. A. S. Kiran Kumar and
Prof. R. Srinivasan and Prof. Ram Sagar (convener). The management board for 1.3 m optical 
telescope consists of Prof. D. Bhattacharya (chairman), Prof. T. Chandrasekhar, Prof. S. K. Ghosh, 
Prof. A. K. Pati, Prof. P. Sreekumar and Dr. Amitesh Omar (convener). Broad vision and much 
needed support to establish  observing facilities of such dimensions from  Prof. K. Kasturirangan
(chairman), Dr. S. D. Sinvhal, Prof. V. S. Ramamurthy and Prof. J. V. Narlikar are thankfully 
acknowledged. Approvals and encouragements to initiate the above projects from all the members 
of the Governing Council of ARIES are gratefully acknowledged. Useful scientific inputs from 
Drs. R.K.S. Yadav, S.B. Pandey, J.C. Pandey, G. Maheswar, Y.C. Joshi, A. C. Gupta, B.J.Medhi, 
S. Joshi and engineering and technical help from J. Pant, Vishal Shukla, S. Yadava, 
Tarun Bangia, K.G. Gupta and Sh Harish Tiwari are thankfully acknowledged.

% references 
%_____________________________________________________________________________

\bibliography{ms}   %>>>> bibliography data in report.bib
\bibliographystyle{spiebib}   %>>>> makes bibtex use spiebib.bst

\end{document}